\documentstyle[12pt]{article}
\newcommand{\bd}{\begin{displaymath}}
\newcommand{\ed}{\end{displaymath}}
\newcommand{\be}{\begin{equation}}
\newcommand{\ee}{\end{equation}}

\title{Fateev-Zamolodchikov spin chain: excitation spectrum,\\completeness and
thermodynamics}
\author{Giuseppe Albertini\thanks{E-mail address:
albert@max.physics.sunysb.edu}\\Institute for Theoretical Physics\\
SUNY at Stony Brook\\ Stony Brook NY 11794 USA}
\date{}

\begin{document}

\maketitle

\begin{abstract}
The sector of zero $Z_{N}$-charge is studied for the ferromagnetic (FM)
and antiferromagnetic (AFM) version of the $Z_{N}\times Z_{2}$ invariant
Fateev-Zamolodchikov quantum spin chain. We conjecture that the relevant
Bethe ansatz equations should admit, beside the usual string-like
solutions, exceptional multiplets, and a number of non-physical solutions.
Once the physical ones are identified, we show how to get completeness and
the gapless excitation spectrum. The central charge is computed from
the specific heat and found to be $c=2\frac{N-1}{N+2}$ (FM) and $c=1$ (AFM).
\end{abstract}

\newpage

\section{Introduction}

It is of interest to study the Fateev-Zamolodchikov spin chain
\be\label{e1}
H=-\sum_{k=1}^{M}\sum_{n=1}^{N-1}\frac{1}{\sin(n\pi/N)}(X_{k}^{n}+
Z_{k}^{n}Z_{k+1}^{-n})
\ee
where the operators $X_{k}$ and $Z_{k}$ act on the $N^{M}$-dimensional vector
space spanned by the basis $\underline{n}=|n_{1},n_{2},\dots n_{M}>$, $n_{k}=
0,1 \dots N-1$
\begin{eqnarray*}
X_{k}|n_{1} \dots n_{k} \dots n_{M}> & = & |n_{1} \dots n_{k}+1 \dots n_{M}>
\ \ {\rm mod} N\\
Z_{k}|n_{1} \dots n_{k} \dots n_{M}> & = & \omega^{n_{k}}|n_{1} \dots n_{k}
\dots n_{M}> \ \ \omega=\exp(2\pi i/N)
\end{eqnarray*}
in its ferromagnetic (i.e. $H$ itself) and its antiferromagnetic version (i.e.
$-H$). The Hamiltonian (\ref{e1}) is exactly integrable since it can be
derived from
the family of commuting transfer matrices $T(u)$ of an integrable
2-dimensional spin model \cite{FZ}
\begin{eqnarray*}
[T(u),T(u')]&=&0 \ \ \  \forall u,u' \in C\\
T(u)=1_{id}&-&Mu\sum_{n=1}^{N-1}\frac{1}{\sin(n\pi/N)}-uH+O(u^{2})
\end{eqnarray*}
As it was conjectured in the paper \cite{ZF}, and subsequently proved
through the calculations of the critical exponents \cite{JMO}, the model is
critical and, in the scaling limit, it gives a $Z_{N}$ invariant conformal
field theory with parafermion currents.

While this leaves little to be discovered about the ferromagnetic (FM)
version of (\ref{e1}), little is still known about the antiferromagnetic
version (AFM). The case $N=3$, corresponding to the three-state
critical Potts chain,
has been studied in great detail, leading to the calculation of the excitation
spectrum \cite{ADM1}, the central charge and the spectrum of conformal
dimensions of the corresponding field theory \cite{Kd,PK} and
finally the characters of the relevant representations of the Virasoro algebra
as well as the full modular invariant partition function \cite{KM}.
On one hand,
these results have led to recognize that the $N=3$ AFM spin chain is critical
and in the scaling limit it describes, rather surprisingly, a conformal
field theory with $Z_{4}$ parafermions. On the other hand, even the results
for $N=3$ FM have proven fruitful because they have provided a new way,
directly related to lattice models, to express characters of the
representations of the Virasoro algebra and affine Lie algebras \cite{KKMM}.

In this paper we extend some of these results to arbitrary (but odd) N. We
summarize here what is known about the exact diagonalization of
(\ref{e1}) \cite{Al}. By
means of the analitic Bethe ansatz, the whole spectrum of (\ref{e1}) can be
expressed in terms of a set of variables $\{\lambda_{k}\}$, related in a
simple
way to the zeroes of the transfer matrix eigenvalues of the two dimensional
model. The
set $\{\lambda_{k}\}$ obeys a system of trascendental equations whose
appearance is commonplace in integrable models
\begin{eqnarray}\label{e2}
&&\prod_{k=1}^{L}\frac{\sinh(\lambda_{j}-\lambda_{k}+i\gamma)}
{\sinh(\lambda_{j}-\lambda_{k}-i\gamma)}=(-)^{M+1}[\frac{\sinh(\lambda_{j}
+is\gamma)}{\sinh(\lambda_{j}-is\gamma)}]^{2M}\\
&&\gamma=\frac{(N-1)\pi}{2N}\ \ \ s=\frac{1}{2(N-1)} \nonumber\\
&&L=(N-1)M-2Q\ \ \ Q=0,1 \dots (N-1)/2 \nonumber
\end{eqnarray}
Here $Q$ is the $Z_{N}$ charge, defined mod$N$, corresponding to the
eigenvalue of the conserved operator
\bd
\exp (2i\pi Q/N)=\prod_{k=1}^{M}X_{k}
\ed
Sectors of charge $Q$ and $N-Q$ are mapped into each other by the (conserved)
charge conjugation operator
\bd
C|n_{1},n_{2},\dots n_{M}>=|N-n_{1},N-n_{2}, \dots N-n_{M}>
\ed
so that the symmetry group of (\ref{e1}) is $Z_{N} \times Z_{2}$.
Energy and momentum of each state are related to a solution of (\ref{e2})
\begin{eqnarray}
E&=&\sum_{k=1}^{L}\cot (i\lambda_{k}+\frac{\pi}{4N})\ \ -2M \!\sum_{k=1}
^{(N-1)/2} \! \cot (\pi k/N) \label{e3} \\
\exp (iP)&=&\prod_{k=1}^{L}\frac{\sinh(\lambda_{k}+\frac{i\pi}{4N})}
{\sinh(\lambda_{k}-\frac{i\pi}{4N})} \label{e4}
\end{eqnarray}
In the following sections we consider the sector $Q=0$ and show that the
excitation spectrum is massless
and made up of one species of ``quasiparticle'' (FM) and $N$ species of
``quasiparticles'' (AFM) with linear dispersion relations at small momenta
\begin{eqnarray}
{\rm FM}&\ \ & \epsilon (p) \approx Np \label{e5} \\
{\rm AFM}&\ \ & \epsilon (p) \approx \frac{N}{N-1}p \label{e6}
\end{eqnarray}
We also determine a set of completeness rules that allow to classify all
physical solutions of (\ref{e2}), i.e those solutions that
actually correspond to
an eigenstate of (\ref{e1}), and show how to count them to obtain the correct
dimension of the sector $Q=0$
in the vector space spanned by the basis $\underline{n}$.
Finally, the set of nonlinear
integral equations \cite{YY} that describe the thermodynamics of (\ref{e1}),
are used to
determine the low temperature asymptotics of the specific heat, related
to the central charge of the conformal field theory by \cite{Af}
\bd
C\approx \frac{\pi c}{3v}T
\ed
where $v$ is the group velocity of massless excitation, from which we find
\begin{eqnarray}
{\rm FM}&\ \ & c=2 \frac{N-1}{N+2} \label{e7} \\
{\rm AFM}&\ \ & c=1 \label{e8}
\end{eqnarray}
The value (\ref{e7}) is of course the one predicted in \cite{ZF}.

\section{Strings and multiplets}

The traditional hypothesis on the solutions of (\ref{e2}) is that, in the limit
$M\rightarrow \infty$, they group into strings
\be\label{e9}
\lambda_{l,\alpha}^{(n,v)}=\lambda_{\alpha}^{(n,v)}+\frac{i\gamma}{2}(n+1-2l)+
\frac{i\pi(1-v)}{4}\ \ \ \ l=1,2 \dots n
\ee
where $\lambda_{\alpha}^{(n,v)}$ is the real center of the string, $n$ its
length, and $v=\pm 1$ its parity. It's been long known \cite{Wy},
and the numerical
analysis of the case $N=3$ confirms it \cite{ADM2}, that complex pairs
with imaginary part different from the one given in (\ref{e9}) can
also appear. We formulate the
following conjecture: for fixed $N=2p+1$, the solutions (roots) of (\ref{e2})
fall into three classes
\begin{enumerate}
\item 1-strings with both parities: $(1,v),\ \ v=\pm1$
\item Even length strings: $(n,v),\ \ n=2,4,\dots N-1,\ \  v=\pm1$
\item $p$ multiplets, to be denoted $(M,m)$, of length $4m+2$,
$m=0,1,\dots p-1$
\end{enumerate}
\be\label{e10}
\lambda_{l,\alpha}^{(M,m)}=\lambda_{\alpha}^{(M,m)}\pm i(\frac{\pi}{4}+\frac
{l\pi}{2N})\ \ \ \ l=-m,-m+1,\dots m-1,m
\ee
with $\lambda_{\alpha}^{(M,m)}$ real. So, for $N=3$ $(p=1)$ we have only a pair
$(m=0)$, for $N=5$ $(p=2)$ we have a pair $(m=0)$ and a sextet $(m=1)$, etc..
This conjecture is partly motivated by a numerical diagonalization of
the transfer matrix and partly warranted {\em a posteriori} by the fact
that it produces the
correct counting of states (see Section 4). To diagonalize $T(u)$ numerically,
one fixes the
spectral parameter $u$ at some conveniently chosen complex value, finds the
eigenvectors numerically and then applies $T(u)$ to them. The eigenvalues are
then polynomials in $e^{2iu}$ and it is easy to locate numerically their
zeroes \cite{ADM2}.

Equations for the real centers are obtained by taking the product of
(\ref{e2})  over
members of a string (multiplet), so that all factors are in the form
\begin{eqnarray*}
&&G(\lambda,\alpha,v)=\frac{\sinh(i\alpha +\lambda +i(1-v)\frac{\pi}{4})}
{\sinh(i\alpha -\lambda +i(1-v)\frac{\pi}{4})}\\
&&\lambda \in R\ \ \alpha \in (-\pi/2,\pi/2] \nonumber
\end{eqnarray*}
In (\ref{e9}) strings have been assigned their limiting ``perfect'' value, and
one may have $\alpha=0$ or $\alpha=\pi/2$ with $G(\lambda,0,v)=-v$ and
$G(\lambda,\pi/2,v)=v$. We group these exceptional factors to yield an
overall $\pm$ sign, and then take the Log choosing the branch
\begin{eqnarray}
&&i{\rm Log}G(\lambda,\alpha,v)\doteq \phi(\lambda,\alpha,v)=
2v\arctan(\cot(\alpha)^{v}\tanh(\lambda)) \label{e11} \\
&& \alpha \in (-\pi/2,0)\cup(0,\pi/2) \nonumber
\end{eqnarray}
so that $\phi(\lambda,\alpha,v)$ is a monotonical continuous function with
values in $(-\pi,\pi)$. The new equations are written with the help of
$Z$-functions
\begin{eqnarray}
Z_{j}(\lambda_{\alpha}^{(j)}) & = & I_{\alpha}^{(j)}/M \label{e12} \\
Z_{j}(\lambda) & \doteq & \frac{1}{2\pi} t_{j}(\lambda)-\frac{1}{2\pi M}
\sum_{k}\sum_{\beta=1}^{M_{k}}\Theta_{j,k}(\lambda-\lambda_{\beta}^{(k)})
\label{e13}
\end{eqnarray}
Here $j,k$ denote the type of string (multiplet), $M_{k}$ is their number,
$I_{\alpha}^{(j)}$ are
integral or half-odd, and the functions in (\ref{e13}) are defined
\begin{eqnarray}
t_{(1,v)}(\lambda)&=&2\phi(\lambda,\frac{\pi}{4N},v) \label{e14} \\
t_{(n,v)}(\lambda)&=&\sum_{l=1}^{n}2\phi(\lambda,\frac{\gamma}{2}
(n+2s+1-2l),v) \label{e15} \\
t_{(M,m)}(\lambda)&=&\sum_{\epsilon=\pm1}2\epsilon
\phi(\lambda,\frac{\pi}{4}+\epsilon \frac{2m+1}{4N},+) \label{e16}
\end{eqnarray}
\begin{eqnarray}
\Theta_{(n,v),(n',w)}(\lambda)&=&\phi(\lambda,\frac{\gamma}{2}(n+n'),vw)
+\phi(\lambda, \frac{\gamma}{2}|n-n'|,vw) \nonumber \\
&+&\sum_{l=1}^{{\rm min}(n,n')-1}2\phi(\lambda,\frac{\gamma}{2}(|n-n'+2l),vw)
\label{e17}
\end{eqnarray}
(including the case $n$ or $n'$ =1)
\be
\Theta_{(1,v),(M,m)}(\lambda)=\sum_{\epsilon=\pm 1}[\epsilon
\phi(\lambda,\frac{\pi}{4}-\frac{\epsilon \pi m}{2N},v)+\epsilon
\phi(\lambda,\frac{\pi}{4}-\frac{\epsilon \pi (m+1)}{2N},v)] \label{e18}
\ee
\begin{eqnarray}
\Theta_{(n,v),(M,m)}(\lambda)&=&\sum_{\epsilon=\pm 1}\sum_{l=1}^{2m}
2\phi(\lambda,\frac{1+\epsilon}{4}\pi-\frac{\pi}{4N}(n-2m-1+2l),v)+
\label{e19} \\
\sum_{\epsilon=\pm 1}[\phi(\lambda,\frac{1+\epsilon}{4}\pi&-&\frac{\pi}{4N}(n-
2m-1),v)+\phi(\lambda,\frac{1+\epsilon}{4}\pi-\frac{\pi}{4N}(n+2m+1),v)]\ \
 \ \ (n>2m) \nonumber \\
\Theta_{(n,v),(M,m)}(\lambda)&=&\sum_{\epsilon=\pm 1}\sum_{l=1}^{n-1}
2\phi(\lambda,\frac{1+\epsilon}{4}\pi-\frac{\pi}{4N}(2m+2l+1-n),v)+
\label{e20} \\
\sum_{\epsilon=\pm 1}[\phi(\lambda,\frac{1+\epsilon}{4}\pi&-&\frac{\pi}{4N}(2m-
n+1),v)+\phi(\lambda,\frac{1+\epsilon}{4}\pi-\frac{\pi}{4N}(2m+n+1),v)]\ \
 \ \ (n\leq 2m) \nonumber
\end{eqnarray}
\begin{eqnarray}
\Theta_{(M,m),(M,m')}(\lambda)&=&2\sum_{\epsilon=\pm 1}
[\phi(\lambda,\frac{1+\epsilon}{4}\pi-\frac{\pi}{2N}|m-m'|,+)
+\phi(\lambda,\frac{1+\epsilon}{4}\pi-\frac{\pi}{2N}(m+m'+1),+)] \nonumber \\
&+&4\sum_{\epsilon=\pm 1}\sum_{l=1}^{2{\rm min}(m,m')}
\phi(\lambda,\frac{1+\epsilon}{4}\pi-\frac{\pi}{2N}(|m-m'|+l),+) \label{e21}
\end{eqnarray}
The notation in (\ref{e14}-\ref{e21}) has been used for the sake of
brevity but it is understood,
in agreememt with the remarks before (\ref{e11}), that in
$\phi(\lambda,\alpha,v)$ $\alpha$ is shifted by periodicity to be
in $(-\frac{\pi}{2},\frac{\pi}{2}]$
and $\phi(\lambda,0,v)=\phi(\lambda,\pi/2,v)=0$. These exceptional values of
$\alpha$ only change the oddness of $I^{(j)}$ in (\ref{e12}) and keeping
track of them one finds
\begin{eqnarray}
I^{(1,+)}&=& {\rm integer(half-odd)\ \ if }\ \ 1+M+M_{(1,+)}+M_{N-1,(-)^{p})}
={\rm even(odd)} \nonumber \\
I^{(1,-)}&=& {\rm integer(half-odd)\ \ if }\ \ 1+M+M_{(1,-)}
+M_{(N-1,(-)^{p+1})}={\rm even(odd)} \nonumber \\
I^{(N-1,(-)^{p+1})}&=& {\rm integer(half-odd)\ \ if }\ \ 1+M_{(1,-)}+
M_{(N-1,(-)^{p+1})}={\rm even(odd)} \nonumber \\
I^{(N-1,(-)^{p})}&=& {\rm integer(half-odd)\ \ if }\ \ 1+M_{(1,+)}+
M_{(N-1,(-)^{p})}={\rm even(odd)} \label{e22} \\
I^{(n,+)}&=& {\rm integer(half-odd)\ \ if }\ \ 1+M_{(n,+)}=
{\rm even(odd)} \nonumber\\
I^{(n,-)}&=& {\rm integer(half-odd)\ \ if }\ \ 1+M_{(n,-)}=
{\rm even(odd)} \nonumber \\
I^{(M,m)}&=& {\rm always\ \ integer} \nonumber
\end{eqnarray}
Energy (\ref{e3}) and momentum (\ref{e4}) become functions of the
string centers. Real part
of the energy and momentum are simply related to the $t$-functions
in (\ref{e14}-\ref{e16})
\begin{eqnarray*}
{\rm Re}e_{j}(\lambda)&=&\frac{1}{4}t'(\lambda)\ \  \forall j \\
p_{j}(\lambda)&=& -\frac{1}{2}t_{j}\ \ \forall j \neq (1,+)\ \ \
({\rm mod}2\pi)\\
p_{(1,+)}(\lambda)&=&-\frac{1}{2}t_{(1,+)}(\lambda)+\pi \ \ \ ({\rm mod}2\pi)
\end{eqnarray*}
but even after summation over string (multiplet) members, the energies retain
an imaginary part
\be\label{e23}
{\rm Im}e_{(n,v)}=-\sum_{k=1}^{n}\frac{\sinh(2\lambda)}{\cosh(2\lambda)-
\cos(\frac{N-1}{2N}\pi(n+1-2k)+\frac{(1-v)}{2}\pi-\frac{\pi}{2N})}
\ee
including $(1,\pm)$-strings and
\be\label{e24}
{\rm Im}e_{(M,m)}=-\sum_{k=-m}^{m}[\frac{\sinh(2\lambda)}{\cosh(2\lambda)+
\sin(\frac{\pi(2k+1)}{2N})}+\frac{\sinh(2\lambda)}{\cosh(2\lambda)+\sin(\frac
{\pi(2k-1)}{2N})}]
\ee
It will be shown, at least in the thermodynamic limit, how these imaginary
parts can be removed assuming suitable correlations between the quantum numbers
$I_{\alpha}^{(j)}$ in (\ref{e12}).

\section{Ground state and excitations. Correlation of rapidities.}

The survival of an imaginary part in the bare energies
(\ref{e23}-\ref{e24}) signals that
correlations between the rapidities $\{\lambda_{\alpha}^{(j)}\}$ must exsist
in order to ensure reality of the total energy. In fact, the detailed
investigation carried out in the simplest $N=3$ case \cite{ADM2}
clearly shows that (\ref{e12}) contain several spurious solutions
and only a subset of the possible
choices of quantum numbers $\{I_{\alpha}^{(j)}\}$ in (\ref{e12})
reproduces the
correct physical solutions, i.e. the spectrum of $H$. This peculiarity,
together with the massive appearance of non-string multiplets, is probably
a consequence of the ``unorthodox'' approach used here to diagonalize the
transfer
matrix \cite{Al}. In fact, the unknowns of (\ref{e2}) are, up to a change
of variables,
the zeroes of the transfer matrix eigenvalues themselves, rather than the
zeroes of an auxiliary $Q$-matrix satisfying a ``$T-Q$ recursion
relation'' \cite{Bx,BR}. Therefore some details of the conventional method of
dealing with (\ref{e12}) must be modified. In particular,
densities of rapidities
and holes that describe the solutions of (\ref{e12}) in the
thermodynamic limit
are expected to be related by a set of constraints \cite{ADM1}
and this should be taken
into account if one were searching for a minimum of the free energy
functional of the gas of rapidities and holes \cite{YY}.

Instead of assuming from the start a set of selection rules on
$\{I_{\alpha}^{(j)}\}$, and consequently on the rapidities
$\{\lambda_{\alpha}^{(j)}\}$, we prefer to make a working assumption,
essentially based on small chain observations, on the ground state in the
infinite chain limit. We will check the validity of the assumption by
performing small variations around the ground state configuration to ensure
that it is indeed a minimum, at least locally, of the energy. The case $N=3$
will provide an heuristic guidance. Since $(N-1,\pm)$ and $(1,\pm)$ strings
play a preminent role in the following, we adopt the shortened,
$N$-dependent notation:
$(1,+)\doteq (a)$, $(1,-)\doteq (b)$, $(N-1,(-)^{p+1})\doteq (c)$,
$(N-1,(-)^{p})\doteq (d)$.

The FM ground state is a band of $(N-1,(-)^{p+1})=(c)$
strings \cite{Al}, whose $I_{\alpha}^{(N-1,(-)^{p+1})}$ form a closely packed
sequence
symmetric around 0, and whose centers fill the real axis with density
\be\label{e25}
\rho_{c}^{(0)}(\lambda)=\frac{2N}{\pi \cosh(2N\lambda)}
\ee
solution of ( here $*$ means convolution) \cite{dV}
\be\label{e26}
\rho_{c}^{0}(\lambda)=-Z'_{c}(\lambda)=-\frac{1}{2\pi}t'_{c}(\lambda)
+\frac{1}{2\pi}\Theta '_{c,c}*\rho_{c}^{(0)}(\lambda)
\ee
The (real) density of ground state energy is
\bd
e_{0}=\lim_{M \rightarrow \infty}\frac{E_{0}}{M}=\int d\lambda \rho_{c}^{0}
(\lambda)e_{c}(\lambda)-2\sum_{k=1}^{(N-1)/2}\cot(\pi k/N)
\ed
In (\ref{e26}) we have taken into account that $Z_{c}$ is a decreasing function
of $\lambda$. Observables in the excited states are computed from dressing
equations \cite{ADM1,BIK} that incorporate the effect of the backflow
in the ground state
distribution (\ref{e25}) when holes are pinched in it and other
types of roots are
added. For a state described by a density of $c$-strings $\rho_{c}$
containing a finite set of $M_{c}^{(h)}$ holes $\{\lambda_{\beta}^{(h,c)}\}$,
and by a finite set
$\{\lambda_{\beta}^{(k)}\}$ of other roots, we have
\bd
Z'_{c}(\lambda)\doteq -\sigma _{c}(\lambda)=-\rho_{c}(\lambda)-
\frac{1}{M}\sum_{\beta =1}^{M_{c}^{(h)}}\delta (\lambda-\lambda_{\beta}^
{(h,c)})
\ed
{}From the definition of $Z$-function
\begin{eqnarray}
\sigma_{c}(\lambda)&-&\frac{1}{2\pi}\Theta'_{c,c}*\sigma_{c}(\lambda)=
-\frac{1}{2\pi}t'_{c}(\lambda) \nonumber \\
&-&\frac{1}{2M\pi}\sum_{\beta=1}^{M_{c}^{(h)}}\Theta'_{c,c}
(\lambda-\lambda_{\beta}^{(h,c)})+\frac{1}{2M\pi}\sum_{k\neq c}
\sum_{\beta=1}^{M_{k}}\Theta'_{c,k}(\lambda-\lambda_{\beta}^{(k)}) \label{e27}
\end{eqnarray}
where the sum $\sum_{k\neq c}$ is taken over all roots other than $c$. Even
without solving (\ref{e27}) explicitly, one sees that the energy
\begin{eqnarray*}
E&=&\sum_{k}\sum_{\beta=1}^{M_{k}}e_{k}(\lambda_{\beta}^{(k)})=\int d\lambda
\sigma_{c}(\lambda)e_{c}(\lambda) \nonumber \\
&-&\sum_{\beta=1}^{M_{c}^{(h)}}e_{c}(\lambda_{\beta}^{(h,c)})+\sum_{k\neq c}
\sum_{\beta=1}^{M_{k}}e_{k}(\lambda_{\beta}^{(k)})
\end{eqnarray*}
can be written as
\be\label{e28}
E=E_{0}-\sum_{\beta=1}^{M_{c}^{(h)}}\epsilon_{c}(\lambda_{\beta}^{(h,c)})+
\sum_{k\neq c}\sum_{\beta=1}^{M_{k}}\epsilon_{k}(\lambda_{\beta}^{(k)})
\ee
where $E_{0}$ is the ground state energy and the dressed energies are
defined as solutions of
\be\label{e29}
\epsilon_{k}-\frac{1}{2\pi}\Theta'_{k,c}*\epsilon_{c}=e_{k}
\ee
In (\ref{e29}) $k$ runs over {\em all} strings and multiplets.
The solution of (\ref{e29})
is easily found by Fourier transform method
\begin{eqnarray}
{\rm Re}\epsilon_{a}(\lambda)&=&-{\rm Re}\epsilon_{c}(\lambda)=
\frac{N}{\cosh(2N\lambda )} \label{e30} \\
{\rm Im}\epsilon_{a}(\lambda)&=&{\rm Im}\epsilon_{c}(\lambda)=
-N\tanh(2N\lambda) \label{e31}
\end{eqnarray}
all other $\epsilon_{k}$ being identically zero. The simplest way to cancel
the imaginary part in (\ref{e28}) is to assume that pairwise
\be
\{\lambda_{\beta}^{(h,c)}\}=\{\lambda_{\beta}^{(a)}\} \ \ \ \beta=1,2 \dots
M_{a} \label{e32}
\ee
Then (\ref{e28}) is non-negative, thereby proving that
(\ref{e25}) is indeed the ground state
distribution (strictly speaking, it has been shown that it is a local
minimum of the energy functional). It would seem that, due to the vanishing
of all other dressed energies, one could add strings (or multiplets) other
than $(a)$ without increasing $E$, but it will be proved in the next section
that this is not the case.

The calculation of momentum is identical. The ground state momentum vanishes
since $\rho_{c}^{(0)}(\lambda)$ is an even function of $\lambda$ and in
an excited state
\be
P=M_{a}\pi +\sum_{k\neq c}\sum_{\beta=1}^{M_{k}}\pi_{k}(\lambda_{\beta}^
{(k)})-\sum_{\beta=1}^{M_{h}}\pi_{c}(\lambda_{\beta}^{(h,c)}) \label{e33}
\ee
where the dressed momenta $\pi_{k}$ solve
\be
\pi_{k}-\frac{1}{2\pi}\Theta'_{k,c}*\pi_{c}=\tilde{p}_{k} \label{e34}
\ee
It is convenient to work with a subtracted (odd in $\lambda$) bare
momentum $\tilde{p}$ defined by $\tilde{p}_{k}=p_{k}$ $\forall k\neq (a)$ and
$p_{a}=\tilde{p}_{a}+\pi$, which accounts for the term $M_{a}\pi$
in (\ref{e33}).
Then $\pi'_{k}=-2{\rm Re}\epsilon_{k}$ which is easily integrated
\bd
\pi_{a}(\lambda)=-\pi_{c}(\lambda)=-2\arctan(\tanh(N\lambda))
\ed
and $\pi_{k}(\lambda)=0$ for all other strings and multiplets.
{}From (\ref{e28}), (\ref{e30})
and (\ref{e32}) it is clear that the excitation spectrum is made
up of one kind of
quasiparticle appearing as a ``bound state'' (not to be understood as a
physical bound state of elementary particles): $(1,+)$ paired to a
hole in the $(N-1,(-)^{p+1})$ band
\begin{eqnarray*}
E&=&E_{0}+\sum_{\beta=1}^{M_{a}}\frac{2N}{\cosh(2N\lambda_{\beta})} \\
P&=&\sum_{\beta=1}^{M_{a}}[\pi-4\arctan(\tanh(N\lambda_{\beta}))]
\end{eqnarray*}
The dispersion curve for each quasiparticle is
\bd
\epsilon(p)=2N\sin(\frac{p}{2}) \ \ \ p\in (0,2\pi)
\ed
which yields (\ref{e5}) at small momenta.

The AFM case has a richer spectrum. We proceed as in the FM case, assuming
a ground state, namely a filled band of $(a)$ and a filled band of $(b)$
corresponding to a closely packed sequence of $I^{(a)}$ and $I^{(b)}$ in
(\ref{e12}). Consequently, their densities are simply the derivatives of the
relevant (increasing) $Z$-functions
\bd
\rho_{j}^{(0)}(\lambda)=Z'_{j}(\lambda)=\frac{1}{2\pi}t'_{j}(\lambda)
-\frac{1}{2\pi}\sum_{l=a,b}\Theta'_{j,l}*\rho_{l}^{(0)}
\ed
where $j$ takes values in $\{a,b\}$. The solution is
\begin{eqnarray}
&&\rho_{a}^{(0)}(\lambda)=\frac{1}{2\pi}\int d\lambda e^{-iq\lambda}
\frac{\sinh(\frac{q\pi}{4})}{\sinh(\frac{q\pi}{2N})\cosh(\frac{q\pi(N-1)}{4N})}
\label{e35} \\
&&\rho_{b}^{(0)}(\lambda)=\frac{1}{2\pi}\int d\lambda e^{-iq\lambda}
\frac{\sinh(\frac{q\pi(N-2)}{4N})}{\sinh(\frac{q\pi}{2N})\cosh(\frac{q\pi(N-1)}
{4N})} \label{e36}
\end{eqnarray}
The ground state energy density is
\be\label{e37}
e_{0}=-\sum_{j=a,b}\int d\lambda \rho_{j}^{(0)}(\lambda)e_{j}(\lambda)+
2\sum_{k=1}^{(N-1)/2}\cot(k\pi/N)
\ee
We perturbe these distributions by pinching two sets of holes
$\{\lambda_{\beta}^{(h,a)}\}$ and $\{\lambda_{\beta}^{(h,b)}\}$ and adding
extra roots of type $k\neq a,b$
\bd
Z'_{j}(\lambda)\doteq \sigma_{j}(\lambda)=\rho_{j}(\lambda)+\frac{1}{M}
\sum_{\beta=1}^{M_{j}^{(h)}}\delta(\lambda-\lambda_{\beta}^{(h,j)})
\ed
where $j\in \{a,b \}$. The following steps are a straightforward
generalization
of the previous case. Densities of vacancies are found from the system of
two integral equations
\begin{eqnarray}
\sigma_{j}(\lambda)&=&\frac{1}{2\pi}t'_{j}(\lambda)-\frac{1}{2\pi}\sum_{l=a,b}
\Theta'_{j,l}*\sigma_{l} \nonumber \\
&+&\frac{1}{2M\pi}\sum_{l=a,b}\sum_{\beta=1}^{M_{l}^{(h)}}\Theta'_{j,l}
(\lambda-\lambda_{\beta}^{(h,l)})-\frac{1}{2M\pi}\sum_{k\neq a,b}
\sum_{\beta=1}^{M_{k}}\Theta'_{j,k}(\lambda-\lambda_{\beta}^{(k)}) \label{e38}
\end{eqnarray}
where $j \in \{a,b \}$. Owing to (\ref{e38}), energy and momentum in the
thermodynamic limit reduce to
\begin{eqnarray}
E&=&E_{0}-\sum_{j=a,b}\sum_{\beta=1}^{M_{j}^{(h)}}\epsilon_{j}(\lambda_{\beta}
^{(h,j)})+\sum_{k\neq a,b}\sum_{\beta=1}^{M_{k}}\epsilon_{k}(\lambda_{\beta}
^{(k)}) \label{e39} \\
P&=&P_{0}+M_{a}\pi-\sum_{j=a,b}\sum_{\beta=1}^{M_{j}^{(h)}}\pi_{j}(\lambda_
{\beta}^{(h,j)})+\sum_{k\neq a,b}\sum_{\beta=1}^{M_{k}}\pi_{k}(\lambda_
{\beta}^{(k)}) \label{e40}
\end{eqnarray}
provided that the dressed observables $\epsilon$ and $\pi$ are defined as
solutions of
\begin{eqnarray}
\epsilon_{k}&+&\frac{1}{2\pi}\sum_{j=a,b}\Theta'_{k,j}*\epsilon_{j}=-e_{k}
\label{e41} \\
\pi_{k}&+&\frac{1}{2\pi}\sum_{j=a,b}\Theta'_{k,j}*\pi_{j}=\tilde{p}_{k}
\label{e42}
\end{eqnarray}
where $k$ runs over all types of roots. The solution reveals that all imaginary
parts are odd functions of $\lambda$ and ${\rm Im}\epsilon_{(M,k)}=0$,
${\rm Im}\epsilon_{(n,+)}=-{\rm Im}\epsilon_{(n,-)}$, ${\rm Im}\epsilon_{a}=
{\rm Im}\epsilon_{c}$, ${\rm Im}\epsilon_{b}={\rm Im}\epsilon_{d}$, suggesting
that, again, a pairing of rapidities must be in effect. One is led to assume
that $M_{(n,+)}=M_{(n,-)}=M_{n}$, $M_{c}=M_{a}^{(h)}$, $M_{d}=M_{b}^{(h)}$
and
\begin{eqnarray}
\{\lambda_{\beta}^{(n,+)}\}=\{\lambda_{\beta}^{(n,-)}\} \ \
\beta=1,2,\dots M_{n} \label{e43} \\
\{\lambda_{\beta}^{(c)}\}=\{\lambda_{\beta}^{(h,a)}\} \ \
\beta=1,2 \dots M_{c} \label{e44} \\
\{\lambda_{\beta}^{(d)}\}=\{\lambda_{\beta}^{(h,b)}\} \ \
\beta=1,2 \dots M_{b} \label{e45}
\end{eqnarray}
As to the real parts we have
\begin{eqnarray*}
{\rm Re}\epsilon_{a}(\lambda)&=&-{\rm Re}\epsilon_{c}(\lambda)=-\frac{N}
{2\cosh(2N\lambda)} \nonumber \\
&-&\frac{2N}{N-1}\sum_{j=0}^{p-1}\frac{\cosh(\frac{2N\lambda}{N-1})
\cos(\frac{(2j+1)\pi}{2(N-1)})}{\cosh(\frac{4N\lambda}{N-1})
+\cos(\frac{(2j+1)\pi}{N-1})}
\end{eqnarray*}
\bd
{\rm Re}\epsilon_{b}(\lambda)=-{\rm Re}\epsilon_{d}(\lambda)={\rm Re}
\epsilon_{a}(\lambda)+\frac{N}{\cosh(2N\lambda)}
\ed
\bd
{\rm Re}\epsilon_{(n,\pm)}(\lambda)=\frac{4N}{N-1}\sum_{j=1}^{n/2}
\frac{\cosh(\frac{2N\lambda}{N-1})\cos(\frac{(n-2j+1)\pi}{2(N-1)})}
{\cosh(\frac{4N\lambda}{N-1})+\cos(\frac{(n-2j+1)\pi}
{N-1})} \ \ 2\leq n\leq N-3
\ed
\bd
{\rm Re}\epsilon_{(M,m)}=\frac{4N}{N-1}\sum_{j=-m}^{m}\frac{\cosh(\frac
{2N\lambda}{N-1})\cos(\frac{j\pi}{N-1})}{\cosh(\frac{4N\lambda}{N-1})
+\cos(\frac{2j\pi}{N-1})}
\ed
It can be checked that ${\rm Re}\epsilon_{j}$ with $j \in \{c,d,(n,v),(M,m) \}$
is positive definite, confirming that the conjecture about the ground state is
correct. Clearly $N$ different massless excitations are present: $p=\frac{N-1}
{2}$ multiplets, plus $\frac{N-3}{2}$ ``bound states'' $\{(n,+),(n,-)\}$
correlated through (\ref{e43}) and finally
the ``bound states'' $\{c,{\rm hole\ \ in}
\ \ a\}$ and $\{d,{\rm hole\ \ in}\ \ b\}$ correlated
through (\ref{e44}-\ref{e45}).
The equations for the momenta are easily integrated noticing that $\pi'_{k}=
2{\rm Re}\epsilon_{k}$
\begin{eqnarray*}
&&\pi_{a}(\lambda)=-\pi_{c}(\lambda)=-\arctan(\tanh(N\lambda))-
\sum_{j=0}^{p-1}\arctan(\frac{\sinh(\frac{2N\lambda}{N-1})}
{\cos(\frac{(2j+1)\pi}{2(N-1)})}) \\
&&\pi_{b}(\lambda)=-\pi_{d}(\lambda)=2\arctan(\tanh(N\lambda))+\pi_{a}
(\lambda) \\
&&\pi_{(n,\pm)}(\lambda)=2\sum_{j=1}^{n/2}\arctan(\frac{\sinh(\frac{2N\lambda}
{N-1})}{\cos(\frac{(n-2j+1)\pi}{2(N-1)})}) \ \ 2\leq n\leq N-3 \\
&&\pi_{(M,m)}(\lambda)=2\sum_{j=-m}^{m}\arctan(\frac{\sinh(\frac{2N\lambda}
{N-1}}{\cos(\frac{j\pi}{N-1})})
\end{eqnarray*}
These expressions for the dressed momenta are not sufficient to fix completely
the dispersion curve for each quasiparticle since we have to deal with the
contribution $\pi M_{a}$ in (\ref{e40}). This will be done in the
next section after
a rule on the string content of each physical solution of (\ref{e2})
is derived.

\section{Completeness. Correlation of quantum numbers.}

The task of counting, let alone finding, all solutions of the system of
equations (\ref{e2}) is very difficult.
Nevertheless, in the framework of the string
hypothesis, one can replace the original equations with (\ref{e12})
and hope that,
to each set of distinct numbers $\{I_{\alpha}^{(j)}\}$ there corresponds
one and only one solution. Counting
solutions becomes then a problem of combinatorics, and the result should
equal the dimension of the vector space on which the Hamiltonian is defined.

The sets $\{I_{\alpha}^{(j)}\}$ cannot be chosen arbitrarily. An obvious
constraint is that the total number of roots in the sector $Q=0$ must be
\be\label{e46}
(N-1)M=M_{a}+M_{b}+(N-1)(M_{c}+M_{d})+\sum_{\pm 1}\sum_{n=2 \atop even}^{N-3}
nM_{(n,v)}+\sum_{m=0}^{p+1}(4m+2)M_{(M,m)}
\ee
because $L=(N-1)M$ in (\ref{e2}). Secondly, since the $Z$-functions
are supposed to
behave monotonically, the $\{I_{\alpha}^{(j)}\}$ must be chosen within the
limits $MZ_{j}(\pm \infty)$. In principle, for a, say, increasing $Z$-function
and a given string content of a state, one should determine the largest
available vacancy $I_{max}$ as the largest integer (or half-odd, according
to (\ref{e22})) {\em strictly} smaller than $MZ_{j}(+\infty)$
and count the integers
(or half-odd) contained in $[-I_{max},I_{max}]$. This gives the set of
available vacancies for $\{I_{\alpha}^{(j)}\}$, that we denote $vac(j)$. It
is a remarkable property of the sector $Q=0$, and this is proven in
Appendix B, that
\be\label{e47}
\frac{vac(j)}{M}=\pm (Z_{j}(+ \infty)-Z_{j}(- \infty))
\ee
where $+(-)$ is used if $Z_{j}$ is increasing (decreasing).

An additional feature of (\ref{e12}) is that, as it was extensively
shown by the
numerical study for the case $N=3$, only a subset of the possible choices
$\{I_{\alpha}^{(j)}\}$, even when constrained by (\ref{e46})
and (\ref{e47}), gives an
eigenvalue of $H$: the string content of an eigenvalue cannot be chosen
arbitrarily and correlations are in effect between the $\{I_{\alpha}^{(j)}\}$.
Motivated by these observations, and by (\ref{e43}-\ref{e45}),
we are led to assume that, for any state
\be\label{e48}
M_{(n,+)}=M_{(n,-)}\doteq M_{n}
\ee
and denoting by $\{I_{\alpha}^{(h,j)}\}$ the empty vacancies (holes) and by
$M_{j}^{(h)}$ their number
\be\label{e49}
M_{a}^{(h)}=M_{c} \ \ \ M_{b}^{(h)}=M_{d}
\ee
Since $vac(j)=M_{j}+M_{j}^{(h)}$, imposing (\ref{e49}) means
\begin{eqnarray*}
Z_{a}(+\infty)-Z_{a}(-\infty)&=&\frac{M_{a}+M_{c}}{M} \\
Z_{b}(+\infty)-Z_{b}(-\infty)&=&\frac{M_{b}+M_{d}}{M}
\end{eqnarray*}
Either one yields, when (\ref{e46}) and (\ref{e47}) are taken into
account, the content rule
\be\label{e50}
(N-2)M_{a}=NM_{b}+2(N-1)M_{d}+\sum_{n=2 \atop even}^{N-3}2nM_{n}+\sum_{m=0}
^{p-1}(4m+2)M_{(M,m)}
\ee
which generalizes the content rule (3.1) of \cite{ADM1}, and it is in
agreement with
all numerical tests performed for $N=5$, $7$. It has to be understood as the
statement that for any eigenvalue the numbers $M_{j}$ must satisfy (\ref{e50})
as well as (\ref{e46}). From (\ref{e46}) and (\ref{e50}) we choose
to take $M_{b}$, $M_{d}$, $M_{n}$ and $M_{(M,m)}$ as independent variables.
We then find, after observing that all $Z$-functions except $Z_{a}$ and
$Z_{b}$ are decreasing
\begin{eqnarray}
&&vac(a)=vac(c)=M+M_{b}+M_{d} \nonumber \\
&&vac(b)=vac(d)=M_{b}+M_{d} \nonumber \\
&&vac(n,+)=vac(n,-)= \frac{2n}{N-2}(M_{b}+M_{d})+\sum_{n'\neq n}(\frac{2nn'}
{N-2}-2{\rm min}(n,n'))M_{n'}  \label{e51} \\
&&+(\frac{2n^2}{N-2}-2n+1)M_{n}+\sum_{m<\frac{n}{2}}(4m+2)(\frac{n}{N-2}-1)
M_{(M,m)}+\sum_{m\geq \frac{n}{2}}2n(\frac{2m+1}{N-2}-1)M_{(M,m)} \nonumber \\
&&vac(M,m)= \frac{4m+2}{N-2}(M_{b}+M_{d})+\sum_{n\leq 2m}n(\frac{4m+2}{N-2}-2)
M_{n}+\sum_{n>2m}(4m+2)(\frac{n}{N-2}-1)M_{n}+ \nonumber \\
&&\sum_{m'\neq m}(\frac{(4m+2)(4m'+2)}{2(N-2)}-4{\rm min}(m,m')-2)M_{(M,m')}+
(\frac{(4m+2)^2}{2(N-2)}-4m-1)M_{(M,m)} \nonumber
\end{eqnarray}
The reduction by $1/2$ of the vacancies of multiplets \cite{ADM2}
has been adopted here too (see Appendix B).
Of course, the number of vacancies must be, by definition, an integer.
The content rule (\ref{e50}), implying that the right hand side
must be divisible
by $N-2$, guarantees that that is the case for $vac(n)$ and $vac(M,k)$.

As to the correlations between $\{I_{\alpha}^{(j)}\}$, from (\ref{e32}) and
(\ref{e44}-\ref{e45}), and
noticing that (\ref{e46}) affects the oddness table (\ref{e22}),
we are led to assume
\begin{eqnarray}
I_{\alpha}^{(a)}&=&-I_{\alpha}^{(h,c)} \ \ \alpha=1,2 \dots M_{a} \label{i1} \\
I_{\alpha}^{(b)}&=&-I_{\alpha}^{(h,d)} \ \ \alpha=1,2 \dots M_{b} \label{i2}
\end{eqnarray}
and, necessarily because of (\ref{e51})
\begin{eqnarray}
I_{\alpha}^{(c)}&=&-I_{\alpha}^{(h,a)} \ \ \alpha=1,2 \dots M_{c} \label{i3} \\
I_{\alpha}^{(d)}&=&-I_{\alpha}^{(h,b)} \ \ \alpha=1,2 \dots M_{d} \label{i4}
\end{eqnarray}
and from (\ref{e43})
\be\label{i5}
I_{\alpha}^{(n,+)}=I_{\alpha}^{(n,-)}
\ee
which generalize in a rather obvious way the selection rules of \cite{ADM2}.
It is
shown in Appendix A that these conditions on the integers do imply the
correlations (\ref{e32}) and (\ref{e43}-\ref{e45}).

The number of physical solutions of (\ref{e12}) is therefore
\be\label{e52}
\sum_{M_{b},M_{d},M_{n},M_{(M,m)}}\left( vac(M_{c}) \atop M_{c} \right)
\left( vac(M_{b}) \atop M_{b} \right) \prod_{n=2 \atop even}^{N-3} \left(
vac(n) \atop M_{n} \right) \prod_{m=0}^{p-1}\left( vac(M,m) \atop M_{(M,m)}
\right)
\ee
where
\bd
M_{c}=M-\frac{1}{N-2}(NM_{d}+2M_{b}+\sum_{n=2 \atop even}^{N-3}2nM_{n}+
\sum_{m=0}^{p-1}(4m+2)M_{(M,m)})
\ed
and the sum is subjected to the constraints
\begin{enumerate}
\item the right hand side of (\ref{e46}) must be divisible by $N-2$
\item $M_{c} \geq 0$
\item $vac(j) \geq M_{j}\ \ \forall j$
\end{enumerate}
The result of (\ref{e52}) is expected to be $N^{M-1}$,
the dimension of the sector
$Q=0$. This has been checked numerically for $N=$7, 9 and it will be proven
analitically for $N=5$. The proof for $N=3$ has already appeared
in \cite{ADM2}.

It should be noticed that the number of vacancies is a piecewise continuous
function of the imaginary parts of the roots belonging to a multiplet and
it exhibits jumps at some special values. The conjecture (\ref{e10}) on
the structure of the multiplets is based on the observation that
it yields the right counting in (\ref{e52}). Therefore, the fact
that (\ref{e52}) give
$N^{M-1}$ should be regarded as a check of the various assumptions made so far.

When $N=5$, (\ref{e46}) reduces to
\bd
M_{a}=(5M_{b}+8M_{d}+4M_{2}+2M_{(M,0)})/3+2M_{(M,1)}
\ed
Hence, there must be an integer $k$, not necessarily positive, such that
\bd
-M_{b}-M_{d}+M_{2}+2M_{(M,0)}=-3k
\ed
which is used to eliminate $M_{2}$ in favor of $k$. Then (\ref{e52}) becomes
\begin{eqnarray}
&&\sum \left( M+M_{b}+M_{d} \atop M-3M_{d}-2M_{b}+2M_{(M,0)}+4k-2M_{(M,1)}
\right) \left( M_{b}+M_{d} \atop M_{d} \right) \label{e53} \\
&&\left( M_{b}+M_{d}+k \atop M_{b}+M_{d}-2M_{(M,0)}-3k \right)
\left( M_{(M,0)}+2k \atop M_{(M,0)}
\right) \left( 2M_{b}+2M_{d}+M_{(M,1)} \atop M_{(M,1)} \right) \nonumber
\end{eqnarray}
where the sum is over $M_{b}$, $M_{d}$, $M_{(M,0)}$, $M_{(M,1)}$, $k$ and
the constraints 2. and 3. apply. We use the integral representation
\begin{eqnarray*}
&&\frac{1}{2\pi i} \oint dz \frac{(1-z^2)^{-1-C}(1+z)^{B}}{z^{1+A}}=\sum_{n=0}
^{2n\leq A}\left( C+n \atop n \right) \left( B \atop A-2n \right) \\
&&\frac{1}{2\pi i} \oint dz \frac{(1+z)^A}{z^{1+B}}=\left( A \atop B \right)
\end{eqnarray*}
to reduce (\ref{e53}) to
\be\label{e54}
(\frac{1}{2\pi i})^{2} \oint dz_{1}dz_{2}S_{1}(z_{1},z_{2})S_{2}(z_{1},z_{2})
\frac{(1+z_{1})^{M}}{z_{2}z_{1}^{M+1}}
\ee
where
\begin{eqnarray*}
S_{1}&=&\sum_{k,M_{(M,0)}\geq 0} \frac{(1+z_{2})^{k}z_{2}^{3k+2M_{(M,0)}}}
{z_{1}^{4k+2M_{(M,0)}}}\left( 2k+M_{(M,0)} \atop M_{(M,0)} \right) \\
S_{2}&=&\sum_{M_{b},M_{d}\geq 0}\left( M_{b}+M_{d} \atop M_{d} \right)
\frac{[(1+z_{2})(1+z_{1})]^{M_{b}+M_{d}}z_{1}^{3M_{d}+2M_{b}}}{(1-z_{1}^{2})
^{2M_{b}+2M_{d}}z_{2}^{M_{b}+M_{d}}}
\end{eqnarray*}
Convergence of all series is guaranteed provided that we take $|z_{1}|\sim
\epsilon^{\alpha}$, $|z_{2}|\sim \epsilon^{\beta}$ with $\epsilon \ll 1$
and $\beta /2 < \alpha < 3\beta/4$. Then one finds
\begin{eqnarray*}
S_{1}(z_{1},z_{2})&=&\frac{z_{1}^{2}(z_{1}^{2}-z_{2}^{2})}{z_{1}^{4}-z_{2}^{3}-
2z_{1}^{2}z_{2}^{2}} \\
S_{2}(z_{1},z_{2})&=&\frac{z_{2}(1-z_{1})^{2}(1+z_{1})}{z_{2}-z_{1}z_{2}
-2z_{2}z_{1}^{2}-z_{1}^{2}-z_{1}^{3}}
\end{eqnarray*}
The integration over $z_{2}$ is performed first. Only the simple pole of
$S_{2}(z_{1},z_{2})$
\bd
z_{2}=\frac{z_{1}^{2}}{1-2z_{1}}
\ed
is encircled by the contour, while the poles of $S_{1}$ lie outside the
path of integration if $\epsilon$ is sufficiently small.
Finally, after some algebra, (\ref{e54}) reduces to
\bd
\frac{1}{2\pi i} \oint dz \frac{(1+z)^{M-1}(1-3z)}{z^{M+1}(1-4z)}=5^{M-1}
\ed
The last equality has been obtained by deforming the contour of integration
around the simple pole at $z=1/4$. The proof given here is just the upgrading
of an old method \cite{Tk}. Unfortunately we haven't been able to
extend the proof to $N>5$. What seems to make higher $N$ harder is not so much
the constraint (\ref{e50}), which can be taken care of by a double contour
integration like for $N=5$, as the complicated form of $vac(n)$ and
$vac(M,m)$.

Clearly, the content rule (\ref{e50}) has some consequences on the
nature of the
excitation spectrum in the thermodynamic limit, as computed in the previous
section. In the FM case, it has been shown that only $\epsilon_{a}$ and
$\epsilon_{c}$ are nonvanishing, but (\ref{e50}) entails that roots other than
$(a)$ can be added only by increasing $M_{a}$ and with it the energy.
Hence, the ground state is not degenerate, and moreover no one particle
excitation can exist (at least in this sector).

As to the AFM case, we solve (\ref{e46}) and (\ref{e50}) for $M_{a}$, $M_{b}$
\begin{eqnarray}
\! \! \!&&\! \! M_{a}=\frac{NM}{2}-\frac{NM_{c}}{2}-\frac{(N-2)M_{d}}{2}-
\sum_{n=2 \atop even}^{N-3}nM_{n}-\sum_{m=0}^{p-1}(2m+1)M_{(M,m)}
\label{e55} \\
\! \! \!&&\! \! M_{b}=\frac{(N-2)M}{2}-\frac{(N-2)M_{c}}{2}-\frac{NM_{d}}{2}
-\sum_{n=2 \atop even}^{N-3}nM_{n}-\sum_{m=0}^{p-1}(2m+1)M_{(M,m)} \label{e56}
\end{eqnarray}
We restrict ourselves to the simpler situation of even number of lattice sites.
Then (\ref{e55}-\ref{e56}) show that multiplets and $n$-strings,
$n<N-1$, can appear as
one-particle excitations, whereas $M_{c}+M_{d}$ is bound to be even. The
ground state has $M_{a}=\frac{NM}{2}$, $M_{b}=\frac{(N-2)M}{2}$, which
exactly fill the available vacancies (\ref{e51}).
Since the distributions (\ref{e35}-\ref{e36}) are
even in $\lambda$, the only contribution to the ground state momentum is
\bd
P_{0}=\pi M_{a}^{(0)}=\frac{\pi NM}{2}
\ed
As to the excited states, $\Delta P$ picks a contribution $(M_{a}-
M_{a}^{(0)})\pi$. Owing to the pairing (\ref{e43}-\ref{e45}) and
the arbitrariness of
$2\pi$ in the definition of momentum, we have the following dispersion
relations for the elementary excitation
\begin{enumerate}
\item ``Bound state'' $(c)$ plus hole in $(a)$
\bd
E(\lambda)=2\epsilon_{c}(\lambda)\ \ P(\lambda)=2\pi_{c}(\lambda)+\frac{N\pi}
{2}\ \ \ \ P\in(0,N\pi)
\ed
\item ``Bound state'' $(d)$ plus hole in $(a)$
\bd
E(\lambda)=2\epsilon_{d}(\lambda)\ \ P(\lambda)=2\pi_{d}(\lambda)+\frac{(N-2)
\pi}{2}\ \ \ \ P\in(0,(N-2)\pi)
\ed
\item $(n,+)$-string paired with $(n,-)$-string, $n=2,4,\dots N-3$
\bd
E(\lambda)=2\epsilon_{(n,\pm)}(\lambda) \ \ P(\lambda)=2\pi_{(n,\pm)}(\lambda)
+n\pi \ \ \ \ P\in(0,2n\pi)
\ed
\item Multiplets $m=0,1,\dots p-1$
\bd
E(\lambda)=\epsilon_{(M,m)}(\lambda)\ \ P(\lambda)=\pi_{(M,m)}(\lambda)+
\pi(2m+1)\ \ \ \ P\in(0,(4m+2)\pi)
\ed
\end{enumerate}
Calling $\overline{P}$ the upper limit of the momentum, we have in all four
cases
\bd
E\approx \frac{N}{N-1}P \ \ P\rightarrow 0^{+}
\ed
and
\bd
E\approx \frac{N}{N-1}(\overline{P}-P) \ \ P\rightarrow \overline{P}^{-}
\ed
so that the group velocity is always $|v|=\frac{N}{N-1}$

\section{Thermodynamics and central charge}

In the thermodynamic limit, the states of the spin chain are described by the
density of rapidities $\rho_{j}$ and the density of holes $\rho_{j}^{(h)}$
(alternatively the density of vacancies $\sigma_{j}=\rho_{j}+\rho_{j}^{(h)}$)
related by
\be\label{e57}
(\pm)^{r(j)}\sigma_{j}=Z'_{j}=\frac{1}{2\pi}t'_{j}-\frac{1}{2\pi}\sum_{k}
\Theta'_{j,k}*\rho_{k}
\ee
The factor $(\pm)^{r(j)}$ has been included because $Z(\lambda)$ can be
increasing or decreasing, so $r(a)=r(b)=0$ and $r(j)=1$ if $j\neq a,b$.
It is well known \cite{YY,BR,Bb} that, at finite temperature,
the equilibrium state
is determined by a system of nonlinear integral equations obtained by
minimizing the free energy functional. The standard method, though, has to be
suitably generalized to the present situation because not all densities
$\rho_{j}$, $\rho_{j}^{(h)}$ are independent. The following discussion is
closely patterned after \cite{Kd} where the $N=3$ case has been dealt with.

We assume that, not only for low lying states over the FM and AFM vacua, but
for all states of the chain, the following constraints hold
\begin{eqnarray}
&& \rho_{c}=\rho_{a}^{(h)}\ \ \ {\rm and} \ \ \ \rho_{d}=\rho_{b}^{(h)}
\label{e58} \\
&& \rho_{a}=\rho_{c}^{(h)}\ \ \ {\rm and} \ \ \ \rho_{b}=\rho_{d}^{(h)}
\label{e59} \\
&& \rho_{(n,+)}=\rho_{(n,-)}\doteq \rho_{n} \label{e60}
\end{eqnarray}
Notice that (\ref{e58}) and (\ref{e59}) imply $\sigma_{c}=\sigma_{a}$ and
$\sigma_{d}=\sigma_{b}$.
It is convenient to treat FM and AFM separately \cite{Kd}. We begin with FM and
eliminate $\rho_{c}$, $\rho_{d}$ from the equations (\ref{e57})
with $j\neq c,d$. We define the Fourier transform by
\bd
\hat{f}(q)=\frac{1}{2\pi}\int d\lambda e^{iq\lambda}f(\lambda)
\ed
and introduce the compact notation $[n]\doteq \sinh(\frac{qn\pi}{4N})$ and
$\{ n \}\doteq \cosh(\frac{qn\pi}{4N})$. We then find
\be\label{e61}
\sigma_{j}=a_{j}-\sum_{k}{}^{\prime}T_{j,k}*\rho_{k}
\ee
where $\sum^{\prime}$ is defined to extend to all types of roots other
than $c,d$ and
\begin{eqnarray*}
&&\hat{a}_{j}(q)=\frac{1}{\{1\}}\delta_{j,a} \\
&&\hat{T}_{j,k}(q)=-\frac{[N-2]}{2\{1\}[N-1]}\ \ \ j=a,b \ \ \ k=a,b \\
&&\hat{T}_{j,n}(q)=-\hat{T}_{n,j}(q)=\frac{[n]}{[N-1]}\ \ \ j=a,b \\
&&\hat{T}_{j,(M,m)}(q)=-\hat{T}_{(M,m),j}(q)=\frac{[2m+1]}{[N-1]}\ \ \ j=a,b \\
&&\hat{T}_{n,n'}(q)=\frac{2\{1\}[{\rm min}(n,n')][N-1-{\rm max}(n,n')]}
{[1][N-1]} \ \ (n\neq n') \\
&&\hat{T}_{n,n}(q)=\frac{[N-n-1][n-1]+[n][N-n-2]}{[1][N-1]} \\
&&\hat{T}_{n,(M,m)}(q)=\hat{T}_{(M,m),n}(q)=\frac{2\{1\}[{\rm min}(n,2m+1)]
[N-1-{\rm max(n,2m+1)}]}{[1][N-1]} \\
&&\hat{T}_{(M,m),(M,m')}(q)=\frac{2\{1\}[2{\rm min}(m,m')+1][N-2-2{\rm max}
(m,m')]}{[1][N-1]} \ \ (m\neq m') \\
&&\hat{T}_{(M,m),(M,m)}(q)=\frac{[2m][N-2m-2]+[2m+1][N-2m-3]}{[1][N-1]}
\end{eqnarray*}
Note that the parities of the $n$-strings have disappeared because we
find $\sigma_{(n,+)}=\sigma_{(n,-)}$ that implies, with (\ref{e60}),
$\rho_{(n,+)}^{(h)}=\rho_{(n,-)}^{(h)}$, so that only one of the two
parities is independent and needs to be kept.
Instead, replacing (\ref{e58}) in (\ref{e57}) with $j=c,d$ yields
$\sigma_{c}=\sigma_{a}$
and $\sigma_{d}=\sigma_{b}$, showing that (\ref{e59}) actually
follows from (\ref{e57}) and (\ref{e58}).

The energy density functional
\bd
E(\rho)=\sum_{j}\int d\lambda \rho_{j}(\lambda)e_{j}(\lambda)
\ed
is cast in the effective form by eliminating $\rho_{c}$ and $\rho_{d}$ by means
of (\ref{e58}) and then using (\ref{e61})
\be\label{e62}
E(\rho)=\pi \sum_{j}{}^{\prime}\int d\lambda \rho_{j}(\lambda)a_{j}(\lambda)
+e_{0}
\ee
where, like in (\ref{e61}), $\sum_{j}^{\prime}$ extends
to $(a),(b),(n),(M,m)$ and
$e_{0}$ is the ground state energy density. Note that the functions $a_{j}$
(actually, in the case at hand, only one is nonvanishing) are closely related
to the dressed energies and that, even in the effective form (\ref{e62}), the
energy functional depends on $\rho_{j}$ only and not on $\rho_{j}^{(h)}$.
This is the reason why we have chosen to eliminate $\rho_{c}$ and $\rho_{d}$.
In the entropy density functional \cite{YY}, the degrees of freedom are also
reduced by (\ref{e58}-\ref{e60}), so the effective expression is
\be\label{e63}
S(\rho)=\sum_{j}{}^{\prime}\int d\lambda [\sigma_{j}\log \sigma_{j}-\rho_{j}
\log \rho_{j}-\rho_{j}^{(h)}\log \rho_{j}^{(h)}]
\ee
The minimum condition for the free energy density functional $F(\rho)=E(\rho)-
TS(\rho)$ is the system of integral equations
\be\label{e64}
\epsilon_{j}=\pi a_{j}+T\sum_{k}{}^{\prime}T_{k,j}*\log(1+e^{-\epsilon_{k}/T})
\ee
where $e^{-\epsilon/T}=\rho/\rho^{(h)}$. The free energy itself is
\be\label{e65}
F=-T\sum_{j}{}^{\prime}\int d\lambda a_{j}(\lambda)\log(1+e^{-\epsilon_{j}/T})
(\lambda)+e_{0}
\ee
It can be seen that the finite temperature dressed energies $\epsilon_{j}$
as defined by (\ref{e64}) reduce, at $T=0$ to the actual energies of the
excitations, rather than to the dressed energies as defined by (\ref{e29}).
This is because the pairing ``$a$-hole in $c$'' has been included in the
calculation imposing the constraint (\ref{e58}-\ref{e59}).
We are particularly interested in the behavior of the specific heat at
$T \rightarrow 0$. The regions at $|\lambda|\sim -\frac{1}{2N}\log T$ give the
leading contribution to the $T\ll 1$ asymptotics. We define shifted functions
$f^{*}(\lambda)=f(\lambda-\frac{1}{2N}\log T)$ and compare (\ref{e61})
and (\ref{e64})
in the limit $T \rightarrow 0$, finding
\begin{eqnarray}
&&\rho_{j}^{*}\simeq (-)^{r(j)+1}\frac{1}{2N\pi}f(\frac{\epsilon_{j}^{*}}{T})
\frac{d\epsilon_{j}^{*}}{d\lambda} \label{e66} \\
&&\rho_{j}^{(h)*}\simeq (-)^{r(j)+1}\frac{1}{2N\pi}
(1-f(\frac{\epsilon_{j}^{*}}{T}))\frac{d\epsilon_{j}^{*}}{d\lambda} \label{e67}
\end{eqnarray}
where $f(x)=(1+e^{x})^{-1}$. These relations can be inserted in the entropy,
so that the leading term at $T\rightarrow 0$ can be written solely in terms of
scaled dressed energies $\phi(\lambda)$
\be
S=\sum_{j}{}^{\prime}\frac{T}{N\pi}(-)^{r(j)}\int_{-\infty}^{+\infty}
d\lambda \frac{d\phi_{j}}{d\lambda}[f(\phi_{j})\log(1+e^{\phi_{j}})+
(1-f(\phi_{j}))\log(1+e^{-\phi_{j}})] \label{e68} \\
\ee
\be\label{e69}
\phi(\lambda)=\lim_{T \rightarrow 0}\frac{1}{T} \epsilon(\lambda-\frac{1}
{2N}\log T)
\ee
{}From their definition and (\ref{e64}), the scaled dressed energies solve
\be\label{e70}
\phi_{j}(\lambda)=4Ne^{-2N\lambda}\delta_{j,a}+\sum_{k}{}^{\prime}T_{k,j}*
\log(1+e^{-\phi_{k}})(\lambda)
\ee
We perform a change of integration variables in (\ref{e68}).
When $j=a,b \ \ \phi_{j}
\rightarrow g(\phi_{j})=(1+e^{-\phi_{j}})^{-1}$, while for $j\neq a,b \ \
\phi_{j} \rightarrow f(\phi_{j})$. Then
\bd
S=-\frac{2T}{N\pi}\{\sum_{j=a,b}[L(g(\phi{j}(+\infty)))-L(g(\phi_{j}
(-\infty)))]+\sum_{j\neq a,b}{}^{\prime}[L(f(\phi_{j}(+\infty)))-L(f(\phi_{j}
(-\infty)))]\}
\ed
where $L(x)$ is the dilogarithmic Rogers function \cite{Lw}
\bd
L(x)=-\frac{1}{2}\int_{0}^{x}dy(\frac{\log y}{1-y}+\frac{\log(1-y)}{y})
\ed
The limiting values $\phi(\pm \infty)$ can be obtained from the system of
nonlinear (ordinary) equations to which (\ref{e70})
reduces when $\lambda \rightarrow \pm \infty$. The solutions are,
at $\lambda \rightarrow + \infty$
\begin{eqnarray*}
&&\phi_{a}(+\infty)=\phi_{b}(+\infty)=-\log(N-1) \\
&&\phi_{n}(+\infty)=\log((n+1)^{2}-1)\\
&&\phi_{(M,m)}(+\infty)=\log((2m+2)^{2}-1) \\
\end{eqnarray*}
and at $\lambda \rightarrow -\infty$
\begin{eqnarray*}
&&\phi_{a}(-\infty)=+\infty \ \ \ \phi_{b}(-\infty)=-\log (\frac{\sin^{2}(\frac
{2\pi}{N+2})}{\sin^{2}(\frac{\pi}{N+2})}-1) \\
&&\phi_{n}(-\infty)=\log (\frac{\sin^{2}(\frac{(n+1)\pi}{N+2})}{\sin^{2}(\frac
{\pi}{N+2})}-1) \\
&&\phi_{(M,m)}(-\infty)=\log(\frac{\sin^{2}(\frac{(2m+2)\pi}{N+2})}{\sin^{2}
(\frac{\pi}{N+2})}-1)
\end{eqnarray*}
By means of the identities
\begin{eqnarray}
&&\sum_{k=1}^{N-2}L(\frac{1}{(k+1)^{2}})+2L(\frac{1}{N})=L(1)=
\frac{\pi^{2}}{6} \label{e71} \\
&&\sum_{k=2}^{N}L(\frac{\sin^{2}(\frac{\pi}{N+2})}{\sin^{2}(\frac{k\pi}{N+2})})
=\frac{2(N-1)}{N+2}L(1) \label{e72}
\end{eqnarray}
the leading term of the specific heat $C=T\frac{\partial S}{\partial T}$
turns out to be
\bd
C=\frac{2T\pi}{3N}(\frac{N-1}{N+2})
\ed
from which we find the central charge (\ref{e7}).

The AFM case differs in the fact that, in order to avoid having hole densities
in the effective energy functional, we prefer to eliminate $\rho_{a}$ and
$\rho_{b}$ from (\ref{e57}). The result has the same form as (\ref{e61})
but now $T_{j,k}$ is symmetric and
\begin{eqnarray*}
&&\hat{a}_{c}(q)=\frac{[N]}{[2]\{N-1\}} \ \ \ \ \hat{a}_{d}(q)=\frac{[N-2]}{[2]
\{N-1\}} \\
&&\hat{a}_{n}(q)=\frac{[n]}{[1]\{N-1\}} \ \ \ \ \hat{a}_{(M,m)}=\frac{[2m+1]}
{[1]\{N-1\}} \\
&&\hat{T}_{c,c}(q)=\hat{T}_{c,d}(q)=\hat{T}_{d,d}(q)=
\frac{[N-2]}{2[1]\{N-1\}} \\
&&\hat{T}_{c,n}(q)=\hat{T}_{d,n}(q)=\frac{\{1\}[n]}{[1]\{N-1\}} \ \ \
\hat{T}_{c,(M,m)}(q)=\hat{T}_{d,(M,m)}(q)=\frac{\{1\}[2m+1]}{[1]\{N-1\}} \\
&&\hat{T}_{n,n'}(q)=\frac{2[{\rm min}(n,n')]\{1\}\{N-1-{\rm max}(n,n')\}}
{[1]\{N-1\}} \ \ (n\neq n') \\
&&\hat{T}_{n,n}(q)=\frac{[n]\{N-n-2\}+[n-1]\{N-n-1\}}{[1]\{N-1\}} \\
&&\hat{T}_{n,(M,m)}(q)=\frac{2[{\rm min}(n,2m+1)]\{1\}\{N-1-
{\rm max}(n,2m+1)\}}{[1]\{N-1\}} \\
&&\hat{T}_{(M,m),(M,m')}(q)=\frac{2[2{\rm min}(m,m')+1]\{1\}\{N-2-2{\rm max}
(m,m')\}}{[1]\{N-1\}} \ \ (m\neq m') \\
&&\hat{T}_{(M,m),(M,m)}(q)=\frac{[2m]\{N-2m-2\}+[2m+1]\{N-2m-3\}}{[1]\{N-1\}}\\
\end{eqnarray*}
As to the equations with $\sigma_{a}$ and $\sigma_{b}$ in (\ref{e57}),
once (\ref{e59}) is
replaced we find $\sigma_{a}=\sigma_{c}$ and $\sigma_{b}=\sigma_{d}$, i.e.
either one of (\ref{e58}) and (\ref{e59}) implies the other when
inserted in (\ref{e57}).
With the new range of $\sum^{\prime}$ and the new definitions of $a_{j}$ and
$T_{j,k}$, (\ref{e62}), (\ref{e63}), (\ref{e64}), and (\ref{e65})
are formally the same
(of course $e_{0}$ in (\ref{e62}) is now the AFM ground state energy density,
as given in (\ref{e37})). Again $T=0$ in (\ref{e64}) yields the energies
of the zero
temperature excitations, i.e. the zero temperature dressed energies after
they have been paired by the correlations (\ref{e43}-\ref{e45}).

In the $T \ll 1$ limit, the relevant region is $|\lambda|\sim -\frac{N-1}{2N}
\log T$, so the shifted functions will read $f^{*}(\lambda)=f(\lambda-
\frac{N-1}{2N}\log T)$ and (\ref{e66}-\ref{e67}) is replaced
by ($f(x)$ is still $(1+e^{x})^{-1}$
\begin{eqnarray}
&&\rho_{j}^{*}\simeq -\frac{N-1}{2N\pi}f(\frac{\epsilon_{j}^{*}}{T})\frac
{d\epsilon_{j}^{*}}{d\lambda} \label{e73} \\
&&\rho_{j}^{(h)*}\simeq -\frac{N-1}{2N\pi}(1-f(\frac{\epsilon_{j}^{*}}{T}))
\frac{d\epsilon_{j}^{*}}{d\lambda} \label{e74}
\end{eqnarray}
Defining scaled dressed energies $\phi(\lambda)$ as in (\ref{e70}),
we see that they take on the values at $\lambda \rightarrow +\infty$
\begin{eqnarray*}
&& \phi_{c}(+\infty)=\phi_{d}(+\infty)=\log(N-1) \\
&& \phi_{n}(+\infty)=\log((n+1)^{2}-1)\\
&&\phi_{(M,m)}(+\infty)=\log((2m+2)^{2}-1)
\end{eqnarray*}
and at $\lambda \rightarrow -\infty$
\bd
\phi_{j}(-\infty)=+\infty \ \ \ \forall j
\ed
Using (\ref{e73}-\ref{e74}), the entropy is now reduced to
\begin{eqnarray*}
&&S=\frac{(N-1)T}{N\pi}\sum_{j}{}{}^{\prime}\int_{-\infty}^{+\infty}d\lambda
\frac{d\phi_{j}}{d\lambda}[f(\phi_{j})\log(f(\phi_{j}))+(1-f(\phi_{j}))
\log(1-f(\phi_{j}))] \\
&&=\frac{2T(N-1)}{N\pi}\sum_{j}{}^{\prime}[L(f(\phi_{j}(+\infty)))-
L(f(\phi(-\infty)))]
\end{eqnarray*}
and the identity (\ref{e71}) is sufficient to determine the specific
heat leading term
\bd
C=\frac{\pi T(N-1)}{3N}
\ed
from which, since the velocity of the excitations is $\frac{N}{N-1}$, $c=1$
follows.

\section*{Acknowledgements}

I would like to thank Dr. S.Dasmahapatra, Dr. R.Kedem, Prof. B.M.McCoy
and Dr. E.Melzer for useful discussions and some valuable advise about
Mathematica programming. Special thanks go to the faculty and staff of
Nordita, where part of this work has been carried out, for their kind
hospitality. This work was partially supported by the National Science
Foundation under grant \# PHY9309888.

\section*{Appendix A}

In this appendix we prove that the rules (\ref{i1}-\ref{i5}) imply
the correlations of
rapidities, at least for finite energy excited states over FM and
AFM vacua. We begin with FM and consider states having an arbitrarily high
but finite number of holes in the $(c)$ distribution and an arbitrarily high
but finite number of other roots. We then have, neglecting terms of
order $O(1/M)$
\begin{eqnarray}
&&Z_{c}(\lambda)=\frac{1}{2\pi}t_{c}(\lambda)-\frac{1}{2\pi}\Theta_{c,c}*
\rho_{c}^{(0)}(\lambda) \label{a1} \\
&&Z_{a}(\lambda)=\frac{1}{2\pi}t_{a}(\lambda)-\frac{1}{2\pi}\Theta_{a,c}*
\rho_{c}^{(0)}(\lambda) \label{a2}
\end{eqnarray}
It's easy to see, from their Fourier transform, that $Z'_{c}(\lambda)=
-Z'_{a}(\lambda)$ and therefore
\bd
Z_{c}(\lambda)=-Z_{a}(\lambda)
\ed
since they are both odd in $\lambda$. Furthermore, (\ref{a1})
and (\ref{a2}) imply that
they are both monotonic. But, by definition,
\bd
Z_{j}(\lambda_{\beta}^{(j)})=\frac{I_{\beta}^{(j)}}{M} \ \ \
Z_{j}(\lambda_{\beta}^{(h,j)})=\frac{I_{\beta}^{(h,j)}}{M}
\ed
so from (\ref{i1}) we have
\be\label{a3}
Z_{a}(\lambda_{\beta}^{(a)})=-Z_{c}(\lambda_{\beta}^{(h,c)})
=Z_{a}(\lambda_{\beta}^{(h,c)})
\ee
and since $Z_{a}$ is monotonic, (\ref{e32}) follows.

In the AFM case, one considers, again discarding terms $O(1/M)$,
\bd
Z_{j}(\lambda)=\frac{1}{2\pi}t_{j}(\lambda)-\frac{1}{2\pi}\Theta_{j,a}*
\rho_{a}^{(0)}(\lambda)-\frac{1}{2\pi}\Theta_{j,b}*\rho_{b}^{(0)}(\lambda)
\ed
with $j=a,b,(n,+),(n,-)$. In this approximation
\begin{eqnarray*}
&&Z_{a}(\lambda)=-Z_{c}(\lambda) \\
&&Z_{b}(\lambda)=-Z_{d}(\lambda) \\
&&Z_{(n,+)}(\lambda)=Z_{(n,-)}(\lambda)
\end{eqnarray*}
and all Z-functions are monotonic, so that, arguing as in (\ref{a3}), the rules
(\ref{i3}-\ref{i5}) imply the pairings (\ref{e43}-\ref{e45}).

\section*{Appendix B}

Let's suppose that the content rule (\ref{e50}) and assumption (\ref{e48})
hold. We want to prove that the number of vacancies as defined by
(\ref{e47}) coincide with the more precise definition. Using (\ref{e46})
and (\ref{e50}) we see that the oddness table (\ref{e22}) is modified
\begin{eqnarray*}
&&I^{(a)},I^{(c)}={\rm integer(half-odd)}\ \ {\rm if}\ \ M+M_{b}+M_{d}=
{\rm odd(even)} \\
&&I^{(b)},I^{(d)}={\rm integer(half-odd)}\ \ {\rm if}\ \ M_{b}+M_{d}=
{\rm odd(even)} \\
&&I^{(n,+)},I^{(n,-)}={\rm integer(half-odd)}\ \ {\rm if}\ \ M_{n}=
{\rm odd(even)}
\end{eqnarray*}
Furthermore, with (\ref{e46}) and (\ref{e50}) we find
\begin{eqnarray*}
&&MZ_{a}(+\infty)=\frac{M+M_{b}+M_{d}}{2} \ \ \ \
MZ_{c}(+\infty)=-\frac{M+M_{b}+M_{d}}{2} \\
&&MZ_{b}(+\infty)=\frac{M_{b}+M_{d}}{2} \ \ \ \
MZ_{d}(+\infty)=-\frac{M_{b}+M_{d}}{2}
\end{eqnarray*}
Consider the case $M+M_{b}+M_{d}=$odd. Then $I^{(a)}$ are integers and
\be\label{a4}
I^{(a)}_{max}=\frac{M+M_{b}+M_{d}-1}{2}
\ee
It is easy to see that the number of integers in $[-I^{(a)}_{max},
I^{(a)}_{max}]$ is $M+M_{b}+M_{d}$. On the other hand, if $M+M_{b}+M_{d}=$
even, $I^{(a)}$ are half-odd. Again (\ref{a4}) holds, and again, the number of
half-odd numbers in $[-I^{(a)}_{max},I^{(a)}_{max}]$ is $M+M_{b}+M_{d}$.
This is exactly the number of vacancies found from $M(Z_{a}(+\infty)-
Z_{a}(-\infty))$. The same argument applies, with some obvious changes,
to $Z_{b},Z_{c},Z_{d}$, noticing that the last two are decreasing and
$I_{max}$ is found from $Z(-\infty)=-Z(+\infty)$.

As to $n$-strings, the result is the same for $(n,+)$ and $(n,-)$ so we
consider only one. We find that the limit at $\lambda \rightarrow -\infty$
can be expressed
\begin{eqnarray*}
&&MZ_{n}(-\infty)=\frac{n(M-M_{c}-M_{d})}{2}-\sum_{n'\neq n}{\rm min}
(n',n)M_{n'}-\frac{(2n-1)M_{n}}{2} \\
&&-\sum_{m}{\rm min}(n,2m+1)M_{(M,m)}=I_{0}^{(n)}+\frac{M_{n}}{2}
\end{eqnarray*}
where $I_{0}^{(n)}$ is an integer. Again, we inspect the two possible cases.
If $M_{n}$ is even, $I^{(n)}$ is half-odd and
\be\label{a5}
I^{(n)}_{max}=I_{0}^{(n)}+\frac{M_{n}-1}{2}
\ee
and the number of vacancies in $[-I_{max},I_{max}]$ is $2I^{(n)}_{0}+M_{n}$,
the same as $M(Z_{n}(-\infty)-Z_{n}(+\infty))$. Likewise, when $M_{n}$ is
odd, $I^{(n)}$ are integers and (\ref{a5}) still holds,
giving the same number of vacancies.

Finally, with multiplets, the $Z$-function is decreasing and from its
definition, (\ref{e46}) and (\ref{e50}) we have
\begin{eqnarray}
&&MZ_{(M,m)}(-\infty)=(2m+1)(M-M_{c}-M_{d})-2\sum_{n}{\rm min}(n,2m+1)
M_{n} \nonumber \\
&&-\sum_{m'\neq m}(4{\rm min}(m,m')+2)M_{(M,m')}-(4m+1)M_{(M,m)} \label{a6}
\end{eqnarray}
{}From (\ref{e55}) and (\ref{e56}) we see that $M-M_{c}-M_{d}$ is
necessarily an even integer, so we can rewrite (\ref{a6}) as
\bd
MZ_{(M,m)}(-\infty)=2I_{0}^{(M,m)}-M_{(M,m)}
\ed
$I_{0}^{(M,m)}$ being integer. As shown in (\ref{e22}), $I^{(M,m)}$ must
always be integers, still in the case $N=3$ \cite{ADM2}, where only pairs
$(m=0)$ are present, it was observed that $I^{(M,0)}$ is even(odd) when
$M_{(M,0)}$ is odd(even). Let's assume this restriction hold for any $N$
and any $m$, and inspect the two possibilities. If $M_{(M,m)}$ is even,
clearly
\be\label{a7}
I_{max}^{(M,m)}=2I_{0}-M_{(M,m)}-1
\ee
Counting the half-odd numbers in $[-I_{max}^{(M,m)},I_{max}^{(M,m)}]$ we find
$2I_{0}^{(M,m)}-M_{(M,m)}$. If $M_{(M,m)}$ is odd, (\ref{a7}) still holds,
but now $I_{max}^{(M,m)}$ is even and the number of even integers in the
interval is the same. We conclude
\bd
vac(M,m)=2I_{0}-M_{(M,m)}=\frac{M}{2}(Z_{(M,m)}(-\infty)-Z_{(M,m)}(+\infty))
\ed
which accounts for the factor $1/2$ used in (\ref{e51})

\end{document}